\documentclass[thmsa]{article}
\usepackage{amsfonts}
\usepackage{amsmath}
\usepackage{amssymb}
\usepackage{amsfonts}
\usepackage{amsmath}
\usepackage{epsfig,multicol}
\usepackage{graphicx}

\input{tcilatex}

\begin{document}

\setcounter{page}{0} \topmargin0pt \oddsidemargin5mm \renewcommand{%
\thefootnote}{\fnsymbol{footnote}} \newpage \setcounter{page}{0} 
\begin{titlepage}
\begin{flushright}
% EMPG\\
\end{flushright}
\vspace{0.5cm}
\begin{center}
{\Large {\bf Turning the Quantum Group Invariant XXZ Spin-Chain Hermitian: A Conjecture on the Invariant Product} }

\vspace{0.8cm}
{ \large Christian Korff}

\vspace{0.5cm}
{\em 
Department of Mathematics, University of Glasgow, \\
University Gardens, Glasgow G12 8QW, UK}
\end{center}
\vspace{0.2cm}
 
\renewcommand{\thefootnote}{\arabic{footnote}}
\setcounter{footnote}{0}

\begin{abstract}
This is a continuation of a previous joint work with Robert Weston on the quantum group invariant XXZ spin-chain (math-ph/0703085). The 
previous results on quasi-Hermiticity of this integrable model are briefly reviewed and then connected with a new 
construction of an inner product with respect to which the Hamiltonian and the representation of the Temperley-Lieb 
algebra become Hermitian.
The approach is purely algebraic, one starts with the definition of a 
positive functional over the Temperley-Lieb algebra whose values can be computed graphically. 
Employing the Gel'fand-Naimark-Segal (GNS) construction for C$^*$-algebras a self-adjoint 
representation of the Temperley-Lieb algebra is constructed when the deformation parameter $q$ lies in a special section of the
unit circle. The main conjecture of the paper is the unitary equivalence of this GNS representation with the 
representation obtained in the previous paper employing the ideas of PT-symmetry and quasi-Hermiticity. An explicit 
example is presented.
\medskip
\par\noindent
\end{abstract}
\vfill{ \hspace*{-9mm}
\begin{tabular}{l}
\rule{6 cm}{0.05 mm}\\
c.korff@maths.gla.ac.uk
\end{tabular}}
\end{titlepage}\newpage 

\section{Introduction}

This is a continuation of a previous paper \cite{KW} where the Hermiticity
properties of the quantum group invariant $XXZ$ spin-chain Hamiltonian \cite%
{Alc87} \cite{PS90} 
\begin{equation}
H=\frac{1}{2}\sum_{n=1}^{N-1}\left\{ \sigma _{n}^{x}\sigma _{n+1}^{x}+\sigma
_{n}^{y}\sigma _{n+1}^{y}+\Delta _{+}~\left( \sigma _{n}^{z}\sigma
_{n+1}^{z}-1\right) \right\} +\Delta _{-}~\frac{\sigma _{1}^{z}-\sigma
_{N}^{z}}{2}  \label{H}
\end{equation}%
have been investigated. Here the anisotropy parameters $\Delta _{\pm }$ are
defined in terms of a single variable $q$,%
\begin{equation}
\Delta _{\pm }=\frac{q\pm q^{-1}}{2}\ ,  \label{Delta}
\end{equation}%
and $\{\sigma _{n}^{x,y,z}\}$ denote the Pauli matrices acting on the $n^{%
\text{th}}$ site of the spin-chain represented by the state space $\mathfrak{%
H}=V^{\otimes N}$ with $V$ being isomorphic to $\mathbb{C}^{2}$. The above
Hamiltonian, besides belonging to an integrable model, is distinguished by
its $U_{q}(sl_{2})$-invariance. Namely, one has the following representation 
$U_{q}(sl_{2})\rightarrow \limfunc{End}\mathfrak{H}$ in terms of the
matrices 
\begin{equation}
q^{\pm S^{z}}=\tprod_{n=1}^{N}q^{\pm \sigma _{n}^{z}/2},\qquad S^{\pm
}=\sum_{n=1}^{N}q^{\frac{\sigma ^{z}}{2}}\otimes \cdots \otimes \underset{n^{%
\text{th}}}{\sigma ^{\pm }}\otimes q^{-\frac{\sigma ^{z}}{2}}\cdots \otimes
q^{-\frac{\sigma ^{z}}{2}},  \label{Urep}
\end{equation}%
which obey the familiar $U_{q}(sl_{2})$-commutation relations%
\begin{equation}
q^{S^{z}}q^{-S^{z}}=q^{-S^{z}}q^{S^{z}}=1,\quad q^{S^{z}}S^{\pm
}q^{-S^{z}}=q^{\pm 1}S^{\pm },\quad \lbrack S^{+},S^{-}]=[2S^{z}]_{q}\ .
\label{QG}
\end{equation}%
Here, as usual, we have set%
\begin{equation*}
\lbrack x]_{q}:=\frac{q^{x}-q^{-x}}{q-q^{-1}}\ .
\end{equation*}

The case when the deformation parameter $q$ lies on the unit circle $\mathbb{%
S}^{1}$ is of particular interest, since then the corresponding lattice
model is believed to correspond in the thermodynamic limit to a CFT with
central charge \cite{Alc87,PS90}%
\begin{equation}
c=1-\frac{6}{(r-1)r},\qquad q=\exp \left( \frac{i\pi }{r}\right) ,\qquad
r\in \mathbb{R}\ .  \label{c}
\end{equation}%
However, for these values of $q$ one easily verifies that $H$ is
non-Hermitian with respect to the conventional scalar product, 
\begin{equation}
q\in \mathbb{S}^{1}:\qquad H\neq H^{\ast }
\end{equation}%
with $\ast $ denoting the Hermitian adjoint (or conjugate transpose in the
case of matrices). It is therefore not clear whether (\ref{H}) constitutes a
well-defined quantum integrable system or is even diagonalizable.
Nevertheless, numerical computations show that the Hamiltonian possesses
real spectrum for $q\in \mathbb{S}^{1}$. This raises the questions whether
the Hamiltonian (\ref{H}) is related to a Hermitian Hamiltonian via
introducing a different inner product or performing a similarity
transformation.

In our previous work we established for the two cases

\begin{enumerate}
\item[(i)] $r=3,4,5,...,N$ and

\item[(ii)] $r\in (N,\infty )\subset \mathbb{R~}$(here $N$ is the number of
sites)
\end{enumerate}

\noindent that the Hamiltonian (\ref{H}) is quasi-Hermitian (see e.g. \cite%
{SGH92,Most04}). That is, we proved by explicit construction that there
exists a \emph{positive, Hermitian and invertible} operator $\eta :\mathfrak{%
H}\rightarrow \mathfrak{H}$ such that%
\begin{equation}
\eta H=H^{\ast }\eta \ .  \label{nHHn}
\end{equation}%
This enabled us to introduce a new inner product 
\begin{equation}
\left\langle \cdot ,\cdot \right\rangle _{\eta }:\mathfrak{H}\times 
\mathfrak{H}\rightarrow \mathbb{C},\qquad \left\langle x,y\right\rangle
_{\eta }:=\left\langle x,\eta y\right\rangle  \label{etaprod}
\end{equation}%
with respect to which the Hamilton operator becomes Hermitian,%
\begin{equation}
\left\langle x,Hy\right\rangle _{\eta }=\left\langle Hx,y\right\rangle
_{\eta },\qquad x,y\in \mathfrak{H}\;.
\end{equation}%
Hence, we rigorously established that (\ref{H}) does indeed give rise to a
well-defined quantum system. There is, however, an important difference
between the cases (i) and (ii): for $r=3,4,5,...,N$ one has first to carry
out a reduction of the state space $\mathfrak{H}\rightarrow \mathfrak{H}%
_{red}$ employing what is known as "quantum group reduction" \cite{RS90} in
the literature in order to remove non-trivial Jordan blocks of the
Hamiltonian. For $r\in (N,\infty )$ such a reduction proved to be
unnecessary.

In both cases, (i) and (ii), we found in \cite{KW} explicit algebraic
expressions for $\eta $ utilising concepts related to $PT$-symmetry (see
e.g. \cite{Bender07,CFMFAF} for recent reviews) and the quantum analogue of
Schur-Weyl duality \cite{Jim86b}. In fact, we established not only
quasi-Hermiticity of the Hamiltonian but of an associated representation of
the Temperley-Lieb algebra $TL_{N}(q)$ which is\emph{\ }obtained from $N-1$\
generators $\{e_{1},e_{2},...,e_{N-1}\}$\ satisfying the commutation
relations%
\begin{eqnarray}
e_{i}^{2} &=&-(q+q^{-1})e_{i},  \notag \\
e_{i}e_{i\pm 1}e_{i} &=&e_{i},  \notag \\
e_{i}e_{j} &=&e_{j}e_{i},\qquad |i-j|>1\;.  \label{TLdef}
\end{eqnarray}%
Again there is a representation $\pi _{TL}:TL_{N}(q)\rightarrow \limfunc{End}%
V^{\otimes N}$ in terms of Pauli matrices by mapping $e_{i}\mapsto \pi
_{TL}(e_{i})=E_{i}$, where the $E_{i}$'s are defined in terms of the local
Hamiltonians%
\begin{equation}
H=\sum_{i=1}^{N-1}E_{i},\qquad E_{i}=\frac{\sigma _{i}^{x}\sigma
_{i+1}^{x}+\sigma _{i}^{y}\sigma _{i+1}^{y}}{2}+\Delta _{+}\frac{\sigma
_{i}^{z}\sigma _{i+1}^{z}-1}{2}+\Delta _{-}\frac{\sigma _{i}^{z}-\sigma
_{i+1}^{z}}{4}\ .  \label{Erep}
\end{equation}%
In \cite{KW} we constructed a quasi-Hermiticity operator $\eta $ such that 
\begin{equation}
\eta E_{i}=E_{i}^{\ast }\eta \ .  \label{etaE}
\end{equation}%
Thus, we obtained a self-adjoint representation of the Temperley-Lieb
algebra $TL_{N}(q)$ with respect to the inner product (\ref{etaprod}). We
briefly review these results in section 2 of this paper. This is one of the
few examples of a non-Hermitian Hamiltonian where quasi-Hermiticity has been
proved by an \emph{explicit} and \emph{exact} construction of the operator $%
\eta $.\smallskip 

The new result in the present article is an entirely independent approach
(which does neither use $PT$-symmetry nor quasi-Hermiticity) to arrive at
the same self-adjoint representation of the Temperley-Lieb algebra for the
case (ii), i.e. when $r\in (N,\infty )$. Namely, we will define for each
fixed spin sector $\mathfrak{H}_{n}$ (here $n$ is the number of down spins)
a positive functional 
\begin{equation*}
\omega _{n}:TL_{N}(q)\rightarrow \mathbb{R},\qquad n=0,1,2,..,N
\end{equation*}%
over the Temperley-Lieb algebra using only its graphical representation in
terms of Kauffman diagrams \cite{Kauf87}. We then employ a well-kown tool
from the representation theory of $C^{\ast }$-algebras, the
Gel'fand-Naimark-Segal (GNS) construction, which provides us with an inner
product $\left\langle \cdot ,\cdot \right\rangle _{\omega _{n}}$ in terms of
the functional $\omega _{n}$ and an associated Hilbert space $\mathfrak{H}%
_{n}^{\text{GNS}}$. The GNS construction is such that the Temperley-Lieb
algebra generators $e_{i}$ act as self-adjoint operators over $\mathfrak{H}%
_{n}^{\text{GNS}}$, thus giving rise to a Hermitian Hamiltonian $%
h=\sum_{i=1}^{N-1}e_{i}$.\smallskip 

The main conjecture of this article is that the two self-adjoint
representations, one induced by $\eta $ and the other by the functionals $%
\{\omega _{n}\}_{n=0}^{N},$ are unitarily equivalent. That is, there exists
for each $n=0,1,...,N$ an isomorphism $U_{n}:\mathfrak{H}_{n}^{\text{GNS}%
}\rightarrow \mathfrak{H}_{n}$ satisfying%
\begin{equation}
U_{n}hU_{n}^{-1}=H,\quad h=\tsum_{i=1}^{N-1}e_{i}
\end{equation}%
and the identity%
\begin{equation}
\left\langle v,w\right\rangle _{\omega _{n}}=\left\langle
U_{n}v,U_{n}w\right\rangle _{\eta },\qquad v,w\in \mathfrak{H}_{n}^{\text{GNS%
}}\ .
\end{equation}%
The left-hand side of the last equality can be computed by purely graphical
means. This provides us with a novel, efficient formalism to investigate the
spectrum, eigenvectors and, in the long term, correlation functions of the
quantum group invariant XXZ Hamiltonian. Moreover, this is also a new
representation theoretic result for the Temperley-Lieb algebra which (to the
best of the author's knowledge) appears not to be contained in the extensive
literature on this subject, see for instance \cite{Jo83,We88,PPM91,Jo99}.
The extension of this graphical calculus to the root of unity case, where a
reduction of the state space needs to be carried out first, is still an open
problem.

\section{Quasi-Hermiticity of the XXZ chain}

In order that the reader can fully appreciate the main conjecture of this
article we briefly review the results obtained for the quasi-Hermiticty
operator $\eta $ in the previous paper \cite{KW}.

\subsection{Discrete symmetries of the Hamiltonian}

We recall the following definitions from \cite{KW}.$\mathfrak{\smallskip }$

\noindent \underline{\textsc{Definition}}.\textsc{\ [parity, time and
spin-reversal].} \emph{Let }$V=\mathbb{C}v_{+1/2}\oplus \mathbb{C}v_{-1/2}$ 
\emph{then we define the linear operator }$P$\emph{\ on }$\mathfrak{H}%
=V^{\otimes N}$ \emph{by setting}%
\begin{equation}
Pv_{\alpha _{1}}\otimes v_{\alpha _{2}}\cdots \otimes v_{\alpha
_{N}}=v_{\alpha _{N}}\otimes v_{\alpha _{N-1}}\cdots \otimes v_{\alpha
_{1}},\quad \alpha _{i}=\pm 1/2\ .  \label{P}
\end{equation}%
\emph{In contrast, the operator }$T$\emph{\ acts on the basis vectors as the
identity,} 
\begin{equation}
Tv_{\alpha _{1}}\otimes \cdots \otimes v_{\alpha _{N}}=v_{\alpha
_{1}}\otimes \cdots \otimes v_{\alpha _{N}},  \label{T}
\end{equation}%
\emph{but is defined to be antilinear, such that} 
\begin{equation}
T\,\lambda v_{\alpha _{1}}\otimes \cdots \otimes v_{\alpha _{N}}=\bar{\lambda%
}v_{\alpha _{1}}\otimes \cdots \otimes v_{\alpha _{N}},\quad \lambda \in 
\mathbb{C}.
\end{equation}%
\emph{Thus, any matrix }$A$\emph{\ (such as the Hamiltonian }$A=H$\emph{) is
transformed into its complex conjugate under the adjoint action of }$T$, $%
TAT=\bar{A}\ .$ \emph{Finally, we introduce the (linear) spin-reversal
operator }$R$\emph{\ by setting}%
\begin{equation*}
Rv_{\alpha _{1}}\otimes \cdots \otimes v_{\alpha _{N}}=v_{-\alpha
_{1}}\otimes \cdots \otimes v_{-\alpha _{N}}\ .
\end{equation*}%
One now easily computes the transformation properties of the quantum group
and Temperley-Lieb generators under the adjoint action of the involutions $%
P,T,R$. They are summarised in Table 1.\medskip 

\begin{center}
\begin{tabular}{||c||c|c|}
\hline\hline
Operator & Temperley-Lieb & Quantum Group \\ \hline\hline
Parity reversal & $PE_{k}P=E_{N-k}^{\ast }$ & $PS^{\pm }P=\left( S^{\mp
}\right) ^{\ast }$ \\ \hline
Time reversal & $TE_{k}T=E_{k}^{\ast }$ & $TS^{\pm }T=\left( S^{\mp }\right)
^{\ast }$ \\ \hline
Spin reversal & $R~E_{k}R=E_{k}^{\ast }$ & $RS^{\pm }R=\left( S^{\pm
}\right) ^{\ast }$ \\ \hline
\end{tabular}%
\medskip

Table 1. Transformations under parity, time and spin reversal.\medskip\ 
\end{center}

\noindent As a trivial consequence of the relation in Table 1, we have the
vanishing commutators%
\begin{equation}
\lbrack PT,H]=[PR,H]=[RT,H]=0
\end{equation}%
for the Hamiltonian $H$. Since $T$ is antilinear and neither $P$ nor $R$ are
positive operators, none of these relations is sufficient to prove
quasi-Hermiticity. Nevertheless, these discrete transformations play an
essential role in the formulation of the quasi-Hermiticity operator $\eta $.

\subsection{Three expressions for the quasi-Hermiticity operator $\protect%
\eta $}

We have the following three equivalent expressions for the quasi-Hermiticity
operator $\eta $. For their derivation and proofs we refer the reader to 
\cite{KW}.

\paragraph{Expression 1: $\protect\eta $ as sum of projectors.}

The first expression for the quasi-Hermiticity operator $\eta $ is closely
related to the idea of quantum group reduction. First one introduces a
special basis, called the path basis, which decomposes the state space with
respect to the action of the quantum group. Broadly speaking one sucessively
"fuses" the spin-1/2 modules $V=V_{1}$ at each lattice site to (higher) spin-%
$j$ modules $V_{2j}$.\smallskip

\noindent \underline{\textsc{Fact}}\textsc{.} The finite-dimensional
irreducible representations of $U_{q}(sl_{2})$ are isomorphic to the
following modules $V_{2j}\cong \mathbb{C}^{2j+1}$ indexed by $j\in \frac{1}{2%
}\mathbb{N}$ and defined through the maps $\pi _{j}:U_{q}(sl_{2})\rightarrow 
\limfunc{End}V_{2j}$ with%
\begin{eqnarray}
\pi _{j}(s^{\pm })\left\vert j,m\right\rangle &=&\sqrt{[j\mp m]_{q}[j\pm
m+1]_{q}}\left\vert j,m\pm 1\right\rangle ,  \notag \\
\pi _{j}(q^{s^{z}})\left\vert j,m\right\rangle &=&q^{m}\left\vert
j,m\right\rangle ,\;\;m=-j,-j+1,...,j-1,j\;.  \label{spin j rep}
\end{eqnarray}

Clearly, there is a choice in which order to execute this "fusing procedure"
and this choice is encoded in a path $\boldsymbol{j}%
=(j_{0}=0,j_{1}=1/2,j_{2},...,j_{N})$. For instance, if $N=3$ we can fuse
the first two sites to a spin-1 or a spin-0 module and then in the next step
obtain either a spin-3/2 or spin-1/2 module yielding the 3 paths 
\begin{equation*}
\boldsymbol{j}=(0,1/2,1,3/2),\ (0,1/2,1,1/2),\ (0,1/2,0,1/2)\ .
\end{equation*}
Here we have rooted each path at $j_{0}=0$ as it is convention in the
literature. The corresponding basis vectors are given explicitly by the
following formulae.\smallskip

Let $\boldsymbol{j}=(j_{0},j_{1},j_{2},...,j_{N})$ be a path on the $sl_{2}$%
-Bratelli diagram, i.e. the set of sequences specified as follows%
\begin{equation}
\Gamma =\{\boldsymbol{j}=(j_{0},j_{1},j_{2}...,j_{N})~|~j_{0}=0,\;j_{k}\geq
0,\;j_{k+1}=j_{k}\pm 1/2\}\ .  \label{bratelli}
\end{equation}%
Then we define for each $\mathbf{\alpha }=(\alpha _{1},...,\alpha
_{N}),\;\alpha _{i}=\pm 1/2$ the vectors%
\begin{equation}
\left\vert \boldsymbol{j},m\right\rangle =\sum_{|\mathbf{\alpha }%
|=m}\left\langle \mathbf{\alpha }|\boldsymbol{j},m\right\rangle ~v_{\alpha
_{1}}\otimes \cdots \otimes v_{\alpha _{N}},\quad
m=-j_{N},-j_{N}+1,...,0,...,j_{N}  \label{path state}
\end{equation}%
with 
\begin{equation}
\left\langle \mathbf{\alpha }|\boldsymbol{j},m\right\rangle
=\prod_{k=1}^{N-1}\left\vert 
\begin{array}{ccc}
j_{k} & \frac{1}{2} & j_{k+1} \\ 
\sum_{i\leq k}\alpha _{i} & \alpha _{k+1} & \sum_{i\leq k+1}\alpha _{i}%
\end{array}%
\right\vert _{q},\qquad m=|\mathbf{\alpha }|=\sum_{k=1}^{N}\alpha _{k}\ .
\label{path matrix}
\end{equation}%
The factors in the product are the Clebsch-Gordan coefficients which are
computed to \cite{KW}%
\begin{equation}
\left\vert 
\begin{array}{ccc}
j & \frac{1}{2} & j+\frac{1}{2} \\ 
m & \alpha & m+\alpha%
\end{array}%
\right\vert _{q}=q^{-\alpha j+\frac{m}{2}}\left( \frac{[j+2\alpha m+1]}{%
[2j+1]}\right) ^{\frac{1}{2}}\   \label{cgc1}
\end{equation}%
and%
\begin{equation}
\left\vert 
\begin{array}{ccc}
j & \frac{1}{2} & j-\frac{1}{2} \\ 
m & \alpha & m+\alpha%
\end{array}%
\right\vert _{q}=2\alpha q^{\alpha (j+1)+\frac{m}{2}}\left( \frac{[j-2\alpha
m]}{[2j+1]}\right) ^{\frac{1}{2}}\;.  \label{cgc2}
\end{equation}%
As long as $q$ is not a root of unity the above basis is well-defined. If $r$
is integer and $3\leq r\leq N$ one has to constrain the set of allowed paths
to the restricted Bratelli diagram 
\begin{equation}
\Gamma ^{(r)}:=\{\boldsymbol{j}\in \Gamma ~|~2j_{k}+1<r,\;k=1,...,N\}\;.
\label{RSOS}
\end{equation}%
Equipped with this particular basis we are in the position to state the
first expression for the quasi-Hermiticity operator $\eta $.\smallskip

\noindent \underline{\textsc{Theorem 1}}.\textsc{\ }\emph{Let }$\left\vert 
\boldsymbol{j},m\right\rangle _{T}$\emph{\ denote the complex conjugate path
basis, i.e. we set}%
\begin{equation}
\left\vert \boldsymbol{j},m\right\rangle _{T}:=T\left\vert \boldsymbol{j}%
,m\right\rangle =\sum_{|\mathbf{\alpha }|=m}\overline{\left\langle \mathbf{%
\alpha }|\boldsymbol{j},m\right\rangle }~v_{\alpha _{1}}\otimes \cdots
\otimes v_{\alpha _{N}},\qquad \boldsymbol{j}\in \Gamma \ .
\label{Tpath state}
\end{equation}%
\emph{Then the following sum over projectors}%
\begin{equation}
\eta =\sum_{\boldsymbol{j},m}\left\vert \boldsymbol{j},m\right\rangle
_{T}~_{^{{}}T}\!\left\langle \boldsymbol{j},m\right\vert ~,  \label{eta}
\end{equation}%
\emph{defines a positive definite, Hermitian and invertible operator which
satisfies (\ref{etaE}). Here }$E_{i}$\emph{\ are the generators of the
Temperley-Lieb algebra in the representation (\ref{Erep}) and the sum over }$%
j$\emph{\ in (\ref{eta}) is restricted to }$\Gamma ^{(r)}$\emph{\ for }$%
3\leq r\leq N$\emph{\ integer and unrestricted for }$r\in (N,\infty )$\emph{.%
}\medskip

Henceforth, it shall always be understood that we take the restricted path
set $\Gamma ^{(r)}$ for the regime (i) and the unrestricted one, $\Gamma ,$
for regime (ii).

\paragraph{Expression 2: $\protect\eta $ in terms of quantum group
generators.}

Given the expression (\ref{eta}) it is natural to ask whether the sum over
the path states can be performed. The answer is positive.\medskip 

\noindent \underline{\textsc{Theorem 2}}\textsc{.} \emph{Let }$\eta
=RC^{\prime }$\emph{\ with }$R$\emph{\ the spin-reversal operator. For fixed 
}$0\leq j\leq N/2$\emph{\ denote by }$\Gamma _{j}$\emph{\ (resp. }$\Gamma
_{j}^{(r)}$\emph{) the subspace spanned by all path vectors with endpoint }$%
j_{N}=j$\emph{\ (i.e. the direct sum of all spin-}$j$\emph{\ modules }$V_{j}$%
\emph{\ obtained from the Clebsch-Gordan decomposition of }$V^{\otimes N}$%
\emph{\ as discussed above). Then the restriction }$C_{j}^{\prime }$\emph{\
of the operator }$C^{\prime }$\emph{\ on this subspace can be expressed in
terms of the quantum group generators as}%
\begin{equation}
C_{j}^{\prime }=(-)^{\frac{N}{2}-j}\sum_{m\in \frac{1}{2}\mathbb{N}}\frac{%
[j-m]_{q}!}{[j+m]_{q}!}~\frac{(S^{-})^{2m}\delta
_{S^{z},m}+(S^{+})^{2m}\delta _{S^{z},-m}}{2^{\delta _{0,m}}}.  \label{C'S}
\end{equation}%
\emph{On the path basis this operator acts as follows}%
\begin{equation}
C^{\prime }\left\vert \boldsymbol{j},m\right\rangle =(-)^{\frac{N}{2}%
-j_{N}}\left\vert \boldsymbol{j},-m\right\rangle ,  \label{C'action}
\end{equation}%
\emph{where }$j_{N}$\emph{\ is the endpoint of the path }$j$\emph{. Thus, we
have in particular that }$C^{\prime 2}=1$\emph{\ or equivalently}%
\begin{equation}
R\eta R=\eta ^{-1}\ .  \label{RetaRinv}
\end{equation}

\paragraph{Expression 3: $\protect\eta $ in terms of the Hecke algebra.}

In order to state the third expression for the quasi-Hermiticity operator we
need another algebra first: the Hecke algebra $H_{N}(q)$\ is generated by $%
N-1$\ letters $\{b_{i}\}_{i=1}^{N-1}$\ obeying the defining relations,%
\begin{equation}
b_{i}b_{i}^{-1}=b_{i}^{-1}b_{i}=1,\quad
b_{i}b_{i+1}b_{i}=b_{i+1}b_{i}b_{i+1},\quad b_{i}b_{j}=b_{j}b_{i},\quad
|i-j|>1  \label{Braid def}
\end{equation}%
and the quadratic relation%
\begin{equation}
(b_{i}+q)(b_{i}-q^{-1})=0\ .  \label{Hecke square}
\end{equation}%
Once more we introduce a representation $\pi _{H}:H_{N}(q)\rightarrow 
\limfunc{End}V^{\otimes N}$\ whose images we denote by capital letters.
Using the homomorphism $\varphi :H_{N}(q)\rightarrow TL_{N}(q)$ with $%
b_{i}\mapsto q^{-1}+e_{i}$ and $b_{i}^{-1}\mapsto q+e_{i}$ we extend the
previously defined representation (\ref{Erep}) of the Temperley-Lieb algebra
to the Hecke algebra by setting%
\begin{equation}
b_{i}\mapsto \pi _{TL}(\varphi (b_{i}))=B_{i}=q^{-1}+E_{i}\ .
\label{Hecke spin}
\end{equation}

\noindent \underline{\textsc{Theorem 3}}\textsc{.} \emph{Set }$\eta =PC$%
\emph{\ with }$P$\emph{\ being the parity operator. Then the restriction }$%
C_{j}$\emph{\ of the operator }$C$\emph{\ to the subspace }$\Gamma _{j}$%
\emph{\ (resp. }$\Gamma _{j}^{(r)}$\emph{) can be expressed in terms of the
Hecke algebra as follows,}%
\begin{equation}
C_{j}=q^{\frac{N(N-4)}{4}+j(j+1)}\mathcal{B}\ ,  \label{Cbeta}
\end{equation}%
\emph{where }$\mathcal{B}$\emph{\ denotes the image of the following special
braid }$\beta $\emph{\ under the representation (\ref{Hecke spin}),}%
\begin{equation}
\beta =\beta _{1}\beta _{2}\cdots \beta _{N-1},\qquad \beta
_{n}=b_{n}b_{n-1}\cdots b_{1}\ .  \label{beta}
\end{equation}%
\emph{Moreover, we have the identities (\ref{Erep}) and} 
\begin{equation}
\lbrack C,C^{\prime }]=0,\text{\qquad }P\eta P=\eta ^{-1}\;.  \label{CC'0}
\end{equation}

\paragraph{Summary of commutation relations.}

The operators detailed in Theorems 1-3 obey the commutation relations
summarized in the table below. \smallskip

\begin{center}
\begin{tabular}{|c|c|c|c|}
\hline\hline
Operator & Hamiltonian & Temperley-Lieb & Quantum Group \\ \hline\hline
$\eta $ & $\eta H=H^{\ast }\eta $ & $\eta E_{k}=E_{k}^{\ast }\eta $ & $\eta
S^{\pm }=S_{\text{op}}^{\pm }\eta $ \\ \hline
$C=P\eta $ & $[C,H]=0$ & $CE_{k}=E_{N-k}C$ & $[C,S^{\pm }]=[C,S^{z}]=0$ \\ 
\hline
$C^{\prime }=R\eta $ & $[C^{\prime },H]=0$ & $[C^{\prime },E_{k}]=0$ & $%
C^{\prime }S^{\pm }=S^{\mp }C^{\prime },\;C^{\prime }S^{z}=-S^{z}C^{\prime }$
\\ \hline
\end{tabular}%
{\small \medskip }

Table 2. Commutation relations for the operator $\eta $ and the two $C$%
-operators.{\small \medskip }
\end{center}

\noindent \underline{\textsc{Remark}}\textsc{.} The construction of the
operators $C$ and $C^{\prime }$ allows one to avoid the summation over the
entire path space in (\ref{eta}) and, thus, is a practical advantage for
explicit computations with the new inner product (\ref{etaprod}). However,
both operators still have an implicit dependence on the decomposition of the
state space $V^{\otimes N}$ into the spin modules $V_{2j}$ through the $j$%
-dependent scalar factors in (\ref{C'S}) and (\ref{Cbeta}). This
decomposition, which is described through the paths in $\Gamma _{j}$ (resp. $%
\Gamma _{j}^{(r)}$), is by no means the only possible way to decompose the
state space $V^{\otimes N}$ into quantum group modules, but represents a
particular choice. Given any other decomposition of $V^{\otimes N}$ (which
might not be necessarily a decomposition into direct sums of modules) these
scalar factors have to be replaced by non-trivial (non-diagonal) matrices
which are not easily computed, even numerically. It is therefore desirable
to find an alternative expression which does not rely on the choice of
decomposition of the state space into quantum group modules. Such a
construction will be presented in the next section.

\section{Graphical calculus \& GNS construction}

From now on we restrict ourselves to the regime (ii), 
\begin{equation*}
q=\exp (i\pi /r),\qquad r\in (N,\infty )\subset \mathbb{R}.
\end{equation*}%
This is the regime where the Hamiltonian has no non-trivial Jordan blocks
and a reduction of the state space is not necessary. In the first part of
this section we define a family of positive functionals over the
Temperley-Lieb algebra in terms of Kauffman diagrams. In the second part we
then employ these functionals to construct self-adjoint representations of $%
TL_{N}(q)$ employing the Gel'fand-Naimark-Segal (GNS) construction. In the
third part we state the main conjecture: the unitary equivalence with the
self-adjoint representation obtained from the operator $\eta $ discussed in
the previous section.

\subsection{Functionals over $TL_{N}(q)$ in terms of oriented Kauffman
diagrams}

We start by adopting the well-known graphical representation of the
Temperley-Lieb algebra generators $e_{i}$ in terms of Kauffman diagrams, see
the graphical depiction below.%
\begin{equation*}
\includegraphics[scale=0.6]{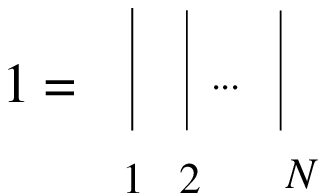}\qquad \text{and}\qquad 
\includegraphics[scale=0.6]{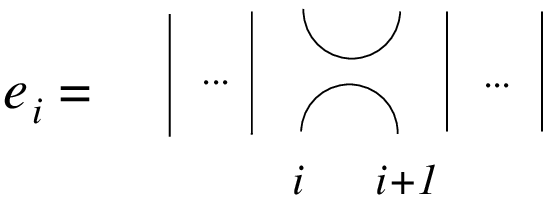}
\end{equation*}%
The Temperley-Lieb algebra acts from above by concatenation of the diagrams.
We now give these diagrams an orientation by introducing "arrow
configurations" such as 
\begin{equation*}
\{\underset{N}{\underbrace{\uparrow \uparrow \downarrow \downarrow \cdots
\uparrow \downarrow \uparrow }}\}\ .
\end{equation*}%
Assigning these arrows (or spins) to the upper and lower ends of the lines
in the Kauffman diagrams, we obtain either clockwise, anti-clockwise or
unoriented cups and caps. Unoriented lines or arcs are those which join
opposing arrows. For instance, the figure below shows two oriented cups (one
anti-clockwise, one clockwise) and two unoriented ones.%
\begin{equation*}
\includegraphics[scale=0.6]{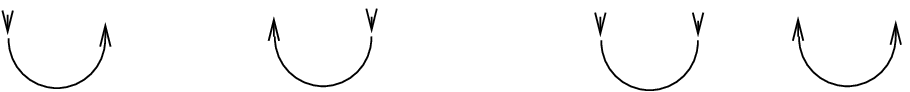}
\end{equation*}%
After these preliminaries we are now ready to state the definition of a
family of functionals.\smallskip 

\noindent \underline{\textsc{Definition}}\textsc{.} \emph{For each integer }$%
0\leq n\leq N$\emph{\ we define the following arrow configuration
(orientation),}%
\begin{equation}
s_{n}=\{\underset{n}{\underbrace{\downarrow \downarrow \cdots \downarrow }}%
\underset{N-n}{\underbrace{\uparrow \uparrow \cdots \uparrow }}\}\ .
\end{equation}%
\emph{Denote by }$t_{a}^{(n)}$\emph{\ the oriented Kauffman diagram
corresponding to }$a\in TL_{N}(q)$\emph{\ with the orientation }$s_{n}$\emph{%
. This diagram will in general contain oriented and\ unoriented cups, caps
and propagating lines as well as closed loops. Denote by }$x_{0}(t_{a}^{(n)})
$\emph{\ the number of unoriented lines and arcs (lines and arcs which join
a pair of arrows pointing in opposite directions), by }$x(t_{a}^{(n)})$\emph{%
\ the number of anti-clockwise oriented cups (concave arcs only) and by }$%
y(t_{a}^{(n)})$\emph{\ the number of closed loops. Then for }$N$\emph{\ even
let }$\omega _{n}:TL_{N}(q)\rightarrow \mathbb{R}$\emph{\ be defined by}%
\begin{equation}
a\mapsto \omega _{n}(a)=\left\{ 
\begin{array}{cc}
0, & \text{if }x_{0}>0 \\ 
(-)^{x+y}(q+q^{-1})^{y}\dfrac{q^{\frac{N}{2}-n}+q^{n-\frac{N}{2}}}{q^{\frac{N%
}{2}-x}+q^{x-\frac{N}{2}}}, & \text{else}%
\end{array}%
\right. \ .  \label{omeven}
\end{equation}%
\emph{For }$N$\emph{\ odd we define instead}%
\begin{equation}
a\mapsto \omega _{n}(a)=\left\{ 
\begin{array}{cc}
0, & \text{if }x_{0}>0 \\ 
(-)^{n+y}(q+q^{-1})^{y}\dfrac{1+\tsum\nolimits_{k=1}^{\frac{N-1}{2}%
-n}(-)^{k}(q^{k}+q^{-k})}{1+\sum_{k=1}^{\frac{N-1}{2}-x}(-)^{k}(q^{k}+q^{-k})%
}, & \text{else}%
\end{array}%
\right. \ .  \label{omodd}
\end{equation}%
\smallskip 

\noindent \underline{\textsc{Examples}}\textsc{.}\ In order to illustrate
the above definition we present two examples, one for $N=4$ and for $N=5$.
Choose the number of down spins to be $n=2$. Then%
\begin{equation*}
N=4:\qquad \omega _{2}(e_{2}^{2}e_{3}e_{1}e_{2})=\includegraphics[scale=0.5]{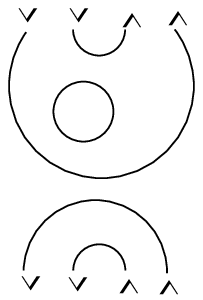}
=-(q+q^{-1}),
\end{equation*}%
since we have $y=1,\;x_{0}=0$ and $x=2$. In contrast, one finds for%
\begin{equation*}
N=5:\qquad \omega _{2}(e_{2}e_{1}e_{3}e_{4}e_{2})=\includegraphics[scale=0.5]{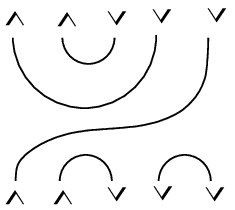}=0
\end{equation*}%
because there is an unoriented propagating line and an unoriented cap, $%
x_{0}>0$.

\subsection{The GNS construction}

First we briefly recall the main facts about the GNS representation. Let $%
\mathfrak{A}$ be a unital $C^{\ast }$-algebra, i.e. $\mathfrak{A}$ is a
complex algebra with unit element $1$ and equipped with a conjugation $\ast :%
\mathfrak{A}\rightarrow \mathfrak{A}$ such that%
\begin{equation*}
(ab)^{\ast }=b^{\ast }a^{\ast },\;(a+b)^{\ast }=a^{\ast }+b^{\ast
},\;(\lambda a)^{\ast }=\bar{\lambda}a^{\ast },\;a^{\ast \ast }=a,\qquad
a,b\in \mathfrak{A},\;\lambda \in \mathbb{C}~.
\end{equation*}%
Let $\omega :\mathfrak{A}\rightarrow \mathbb{C}$ be a "state" over this
algebra, i.e. $\omega $ is a positive linear functional of norm 1,%
\begin{equation}
\omega (a^{\ast }a)\geq 0,~a\in \mathfrak{A}\text{\qquad and\qquad }\omega
(1)=1.
\end{equation}%
There is a natural action of the algebra on itself (seen as vector space) by
assigning to each element $a\in \mathfrak{A}$ the map $\pi _{\omega }(a):%
\mathfrak{A}\rightarrow \mathfrak{A}$ defined as%
\begin{equation*}
b\mapsto \pi _{\omega }(a)b=ab,\qquad b\in \mathfrak{A}\ .
\end{equation*}%
Furthermore, we can endow the algebra with an inner product setting%
\begin{equation}
\langle a,b\rangle _{\omega }:=\omega (a^{\ast }b)\ .
\end{equation}%
In order for this product to be well defined we need to set all elements to
zero for which the induced norm vanishes, i.e. we need to take the vector
space quotient $\mathfrak{A}/\mathfrak{I}$ with respect to the left ideal 
\begin{equation}
\mathfrak{I}=\{a\in \mathfrak{A~}|~\omega (a^{\ast }a)=0\}\ .
\end{equation}%
Finally, taking the norm completion $\mathfrak{H}^{GNS}=\overline{\mathfrak{A%
}/\mathfrak{I}}$ we obtain a Hilbert space and by construction the resulting
representation of $\mathfrak{A}$ over the Hilbert space $\mathfrak{H}^{GNS}$
preserves the $\ast $-structure, i.e. the $\ast $-operation in the algebra
corresponds to taking the Hermitian adjoint with respect to $\langle \cdot
,\cdot \rangle _{\omega }$. Note also that the representation $\pi _{\omega }
$ is cyclic. That is, there exists a vector $\Omega $ (in the present case
the equivalence class of $1\in \mathfrak{A})$ such that $\mathfrak{H}^{GNS}=%
\mathfrak{A}\Omega $ .

We now set $\mathfrak{A}=TL_{N}(q)$ and fix the $\ast $-operation on the
Temperley-Lieb algebra by requiring that $e_{i}^{\ast }=e_{i}$. In terms of
the corresponding Kauffman diagram this operation corresponds to
horizontally flipping the diagram. Then $TL_{N}(q)$ can be turned into a
well-defined $C^{\ast }$-algebra, see e.g. \cite{Jo99}. Following the
general outline just given we obtain for each of the above defined
functionals $\omega _{n}$ a representation $\pi _{n}^{GNS}$ over a Hilbert
space $\mathfrak{H}_{n}^{GNS}$ with inner product%
\begin{equation}
\langle a,b\rangle _{\omega _{n}}:=\omega _{n}(a^{\ast }b)\ .
\end{equation}%
By definition we have that%
\begin{equation}
\langle e_{i}a,b\rangle _{\omega _{n}}=\omega _{n}(a^{\ast }e_{i}b)=\langle
a,e_{i}b\rangle _{\omega _{n}},
\end{equation}%
whence the GNS representation $\pi _{n}^{GNS}$ is self-adjoint, i.e. the
inner product is invariant under the $TL_{N}(q)$ action.{\small \medskip }

\noindent \underline{\textsc{Remark}}\textsc{.} In order to show that this
construction is indeed well-defined one needs to show that the functionals
introduced above are normalized and positive. The correct normalization is
easily verified, positivity on the other hand is more difficult to show and
remains at present a conjecture. However, it has been tested numerically for
many examples, $N=3,4,...,8$. Note, that positivity would be a direct
consequence of the identification with the $\eta $-product (\ref{etaprod})
which we discuss next and which also has been tested numerically.

\subsection{The Conjecture}

Denote by $\mathfrak{H}_{n}\subset \mathfrak{H}=V^{\otimes N}$ the subspace
containing all vectors with $n$ down spins,%
\begin{equation}
\mathfrak{H}_{n}=\limfunc{span}\left\{ v_{\alpha _{1}}\otimes \cdots \otimes
v_{\alpha _{N}}~|~\tsum \alpha _{i}=N/2-n\right\} \ .
\end{equation}%
As it turns out these spin-sectors viewed as Temperley-Lieb modules with
respect to the invariant product $\langle \cdot ,\cdot \rangle _{\eta }$ are
in one-to-one correspondence with the GNS modules.\smallskip

\noindent \underline{\textsc{Conjecture}}\textsc{.} \emph{The self-adjoint
representation }$\pi _{TL}$\emph{\ of the Temperley-Lieb algebra }$TL_{N}(q)$%
\emph{\ with }$q=\exp (i\pi /r),$\emph{\ }$r>N$\emph{\ over the Hilbert space%
} $\{\mathfrak{H}=V^{\otimes N},\langle \cdot ,\cdot \rangle _{\eta }\}$ 
\emph{is unitarily equivalent to the direct sum }$\tbigoplus_{n}\pi
_{n}^{GNS}$\emph{\ of representations over the Hilbert spaces} $\{\mathfrak{H%
}_{n}^{GNS},\langle \cdot ,\cdot \rangle _{\omega _{n}}\}$\emph{. For each
fixed spin sector} $\mathfrak{H}_{n}\subset \mathfrak{H}=V^{\otimes N}$ 
\emph{the unitary map} 
\begin{equation}
U_{n}:\mathfrak{H}_{n}^{GNS}\rightarrow \mathfrak{H}_{n}
\end{equation}%
\emph{is given by}%
\begin{equation}
a\mapsto U_{n}a=\pi _{TL}(a)\Omega _{n},\quad \;\Omega _{n}=\underset{n}{%
\underbrace{v_{-\frac{1}{2}}\otimes \cdots \otimes v_{-\frac{1}{2}}}}\otimes
v_{\frac{1}{2}}\cdots \otimes v_{\frac{1}{2}}\;.
\end{equation}%
\emph{By abuse of notation we do not distinguish between an algebra element }%
$a\in TL_{N}(q)$\emph{\ and its equivalence class with respect to the ideal} 
$\mathfrak{I}_{n}=\{a\mathfrak{~}|~\omega _{n}(a^{\ast }a)=0\}$. \emph{%
Unitarity means that we have the identity}%
\begin{equation}
\omega _{n}(a)=\langle \Omega _{n},\pi _{TL}(a)\Omega _{n}\rangle _{\eta
}\quad \text{\emph{for all}\quad }a\in TL_{N}(q)\;.
\end{equation}%
\smallskip 

\noindent \underline{\textsc{Remark}}\textsc{.} Note that the conjectured
identity can also be seen as a \emph{definition} of the inner product $%
\langle \cdot ,\cdot \rangle _{\eta }$ on $V^{\otimes N}$ via the GNS
construction. Namely, for any basis $\{a_{i}\}\subset TL_{N}(q)$ in the GNS
module $\mathfrak{H}_{n}^{GNS}$ we obtain a basis $\{x_{i}=U_{n}a_{i}=\pi
_{TL}(a_{i})\Omega _{n}\}\subset \mathfrak{H}_{n}$ and vice versa. The
matrix elements of $\eta $ with respect to the basis $\{x_{i}\}$ are then
simply obtained from the Gram matrix of the basis $\{a_{i}\}$ via the
relation%
\begin{equation}
\langle x_{i},\eta x_{j}\rangle =\omega _{n}(a_{i}^{\ast }a_{j})\ .
\label{etaom}
\end{equation}%
This implicit way of introducing $\eta $ has the advantage of being a
basis-independent definition of the new inner product and we have now
removed any dependence on the decomposition of the state space with respect
to the quantum group action. Moreover, this definition of $\eta $ provides
us with an efficient graphical calculus to compute matrix elements with
respect to $\langle \cdot ,\cdot \rangle _{\eta }$. This abstract algebraic
reformulation of our original problem might be of help in extending our
present results to the wider class of physical systems which have
Temperley-Lieb type Hamiltonians.\smallskip

\subsubsection{Example $N=5$}

Let us consider for $N=5$ the sector $S^{z}=1/2,n=2$. Then the following
table gives the correspondence between Kauffman diagrams and basis vectors
(algebra elements),\smallskip

\begin{center}
\begin{tabular}{||l||c|c|c|c|c|c|}
\hline
$a\in TL_{5}$ & $a_{1}=1$ & $a_{2}=e_{2}$ & $a_{3}=e_{1}e_{2}$ & $%
a_{4}=e_{3}e_{2}$ & $a_{5}=e_{4}e_{3}e_{2}$ & $a_{6}=e_{1}e_{3}e_{2}$ \\ 
\hline
diagram & $\includegraphics[scale=0.5]{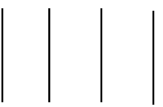}$ & 
$\includegraphics[scale=0.5]{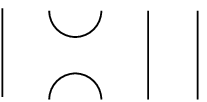}$ & 
\includegraphics[scale=0.5]{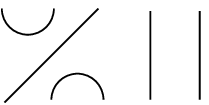} & 
\includegraphics[scale=0.5]{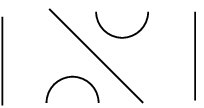} & 
\includegraphics[scale=0.5]{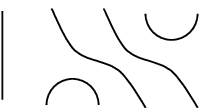} & 
\includegraphics[scale=0.5]{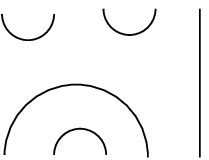} \\ \hline
\end{tabular}
\end{center}

and

\begin{center}
\begin{tabular}{||l|c|c|c|c|}
\hline
$a\in TL_{5}$ & $a_{7}=e_{2}e_{1}e_{3}e_{2}$ & $a_{8}=e_{1}e_{4}e_{3}e_{2}$
& $a_{9}=e_{2}e_{1}e_{4}e_{3}e_{2}$ & $a_{10}=e_{3}e_{2}e_{1}e_{4}e_{3}e_{2}$
\\ \hline
diagram & $\includegraphics[scale=0.5]{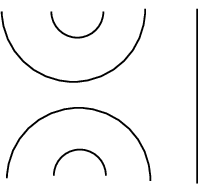}$ & 
\includegraphics[scale=0.5]{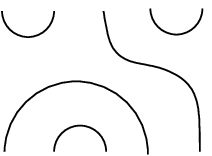} & 
\includegraphics[scale=0.5]{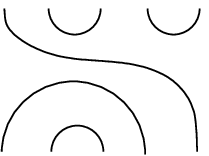} & 
\includegraphics[scale=0.5]{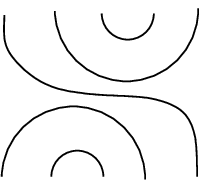}\\ \hline
\end{tabular}%
\smallskip
\end{center}

\noindent We now state the results for computing the Gram matrix $G=(\omega
_{2}(a_{i}^{\ast }a_{j}))_{i,j}$ for this particular basis. From the latter
one can compute $\eta $ via (\ref{etaom}). Denote by $x_{ij}$ the number of
anticlockwise arcs in the diagram associated with $a_{i}^{\ast }a_{j}$ for $%
x_{0}=0$ (no unoriented arcs). Likewise, let $y_{ij}$ be the number of
closed loops asoociated with $a_{i}^{\ast }a_{j}$ for $x_{0}=0$. Both
matrices are symmetric, $x_{ij}=x_{ji}$ and $y_{ij}=y_{ji}$. The values of
the matrix elements $x_{ij},y_{ij}$ are given in the table below, whenever $%
x_{0}\neq 0$ we omit the values from the table. Inserting these values into (%
\ref{omodd}) one obtains the Gram matrix $G$.\medskip

\begin{center}
\begin{tabular}{||l||l|l|l|l|l|l|l|l|l|l|}
\hline\hline
${\small x}_{ij}{\small /y}_{ij}$ & {\small 1} & {\small 2} & {\small 3} & 
{\small 4} & {\small 5} & {\small 6} & {\small 7} & {\small 8} & {\small 9}
& {\small 10} \\ \hline\hline
{\small 1} & \multicolumn{1}{||c|}{\small 0/0} & \multicolumn{1}{|c|}{\small %
1/0} & \multicolumn{1}{|c|}{\small -} & \multicolumn{1}{|c|}{\small -} & 
\multicolumn{1}{|c|}{\small -} & \multicolumn{1}{|c|}{\small -} & 
\multicolumn{1}{|c|}{\small 2/0} & \multicolumn{1}{|c|}{\small -} & 
\multicolumn{1}{|c|}{\small -} & \multicolumn{1}{|c|}{\small -} \\ \hline
{\small 2} & \multicolumn{1}{||c|}{} & \multicolumn{1}{|c|}{\small 1/1} & 
\multicolumn{1}{|c|}{\small 1/0} & \multicolumn{1}{|c|}{\small 1/0} & 
\multicolumn{1}{|c|}{\small -} & \multicolumn{1}{|c|}{\small 2/0} & 
\multicolumn{1}{|c|}{\small 2/1} & \multicolumn{1}{|c|}{\small -} & 
\multicolumn{1}{|c|}{\small -} & \multicolumn{1}{|c|}{\small -} \\ \hline
{\small 3} & \multicolumn{1}{||c|}{} & \multicolumn{1}{|c|}{} & 
\multicolumn{1}{|c|}{\small 1/1} & \multicolumn{1}{|c|}{\small 2/0} & 
\multicolumn{1}{|c|}{\small -} & \multicolumn{1}{|c|}{\small 2/1} & 
\multicolumn{1}{|c|}{\small 2/0} & \multicolumn{1}{|c|}{\small -} & 
\multicolumn{1}{|c|}{\small -} & \multicolumn{1}{|c|}{\small 2/0} \\ \hline
{\small 4} & \multicolumn{1}{||c|}{} & \multicolumn{1}{|c|}{} & 
\multicolumn{1}{|c|}{} & \multicolumn{1}{|c|}{\small 1/1} & 
\multicolumn{1}{|c|}{\small 1/0} & \multicolumn{1}{|c|}{\small 2/1} & 
\multicolumn{1}{|c|}{\small 2/0} & \multicolumn{1}{|c|}{\small 2/0} & 
\multicolumn{1}{|c|}{\small -} & \multicolumn{1}{|c|}{\small -} \\ \hline
{\small 5} & \multicolumn{1}{||c|}{} & \multicolumn{1}{|c|}{} & 
\multicolumn{1}{|c|}{} & \multicolumn{1}{|c|}{} & \multicolumn{1}{|c|}%
{\small 1/1} & \multicolumn{1}{|c|}{\small 2/0} & \multicolumn{1}{|c|}%
{\small -} & \multicolumn{1}{|c|}{\small 2/1} & \multicolumn{1}{|c|}{\small -%
} & \multicolumn{1}{|c|}{\small -} \\ \hline
{\small 6} & \multicolumn{1}{||c|}{} & \multicolumn{1}{|c|}{} & 
\multicolumn{1}{|c|}{} & \multicolumn{1}{|c|}{} & \multicolumn{1}{|c|}{} & 
\multicolumn{1}{|c|}{\small 2/2} & \multicolumn{1}{|c|}{\small 2/1} & 
\multicolumn{1}{|c|}{\small 2/1} & \multicolumn{1}{|c|}{\small 2/0} & 
\multicolumn{1}{|c|}{\small 2/1} \\ \hline
{\small 7} & \multicolumn{1}{||c|}{} & \multicolumn{1}{|c|}{} & 
\multicolumn{1}{|c|}{} & \multicolumn{1}{|c|}{} & \multicolumn{1}{|c|}{} & 
\multicolumn{1}{|c|}{} & \multicolumn{1}{|c|}{\small 2/2} & 
\multicolumn{1}{|c|}{\small 2/0} & \multicolumn{1}{|c|}{\small 2/1} & 
\multicolumn{1}{|c|}{\small 2/0} \\ \hline
{\small 8} & \multicolumn{1}{||c|}{} & \multicolumn{1}{|c|}{} & 
\multicolumn{1}{|c|}{} & \multicolumn{1}{|c|}{} & \multicolumn{1}{|c|}{} & 
\multicolumn{1}{|c|}{} & \multicolumn{1}{|c|}{} & \multicolumn{1}{|c|}%
{\small 2/2} & \multicolumn{1}{|c|}{\small 2/1} & \multicolumn{1}{|c|}%
{\small 2/0} \\ \hline
{\small 9} & \multicolumn{1}{||c|}{} & \multicolumn{1}{|c|}{} & 
\multicolumn{1}{|c|}{} & \multicolumn{1}{|c|}{} & \multicolumn{1}{|c|}{} & 
\multicolumn{1}{|c|}{} & \multicolumn{1}{|c|}{} & \multicolumn{1}{|c|}{} & 
\multicolumn{1}{|c|}{\small 2/2} & \multicolumn{1}{|c|}{\small 2/1} \\ \hline
{\small 10} & \multicolumn{1}{||c|}{} & \multicolumn{1}{|c|}{} & 
\multicolumn{1}{|c|}{} & \multicolumn{1}{|c|}{} & \multicolumn{1}{|c|}{} & 
\multicolumn{1}{|c|}{} & \multicolumn{1}{|c|}{} & \multicolumn{1}{|c|}{} & 
\multicolumn{1}{|c|}{} & \multicolumn{1}{|c|}{\small 2/2} \\ \hline
\end{tabular}%
\medskip 
\end{center}

\noindent Again an illustrative example of how to obtain the above values
might be helpful. Let us consider the matrix element $G_{6,10}=\omega
_{2}(a_{6}^{\ast }a_{10})$. The Kauffman diagram is obtained by flipping the
one for $a_{6}$ at the horizontal axis and then connecting it with the one
for $a_{10}$ from above,%
\begin{equation*}
a_{6}^{\ast }a_{10}=e_{2}e_{3}e_{1}~a_{10}=\includegraphics[scale=0.5]{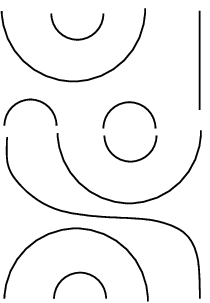}\ .
\end{equation*}%
Obviously, we have $y_{6,10}=1$. Adding the orientation $\{\downarrow
\downarrow \uparrow \uparrow \uparrow \}$ on the top and bottom of the
diagram we see that there are no unoriented lines or arcs, $x_{0}=0$, and
that we have two anti-clockwise oriented cups, $x_{6,10}=2$. Hence, $\omega
_{2}(a_{6}^{\ast }a_{10})=-(q+q^{-1})\ .$

The reader might wonder why we picked out this particular basis among other
choices. The above basis vectors transform particularly simple under the
action of the Temperley-Lieb algebra: the only coefficients which occur in
the expansion are 0, $1$ and powers of $-(q+q^{-1})$. It is in this basis
that $\eta $ takes its simplest form and which is most suitable for the
diagonalization of the Hamiltonian.

\section{Conclusions}

In this article we have presented a conjecture for a new construction of a
self-adjoint representation of the Temperley-Lieb algebra $TL_{N}(q)$ with
deformation parameter $q=\exp (i\pi /r),\;N<r<\infty $. Clearly, in the
thermodynamic limit $N\rightarrow \infty $ this section of the unit circle
shrinks to the point $q=1$ where the Hamiltonian is Hermitian and its
algebraic structure drastically simplifies.

Nevertheless, this construction is significant as it provides a practical
advantage in computing the quasi-Hermiticity operator $\eta $ for finite
lattice size and, as for now, is the simplest starting point for the
computation of the square root $\eta ^{1/2}$. The latter is needed to obtain
the Hamiltonian%
\begin{equation}
h=\eta ^{1/2}H\eta ^{-1/2},  \label{littleh}
\end{equation}%
which is Hermitian with respect to the original scalar product on the state
space. This Hamiltonian might look very different form the initial
Hamiltonian (\ref{H}) and there is a priori no reason to expect that it will
have only nearest neighbour bulk interaction as the similarity
transformation is highly nonlocal. Preliminary computations for the case
when $q=\exp (i\pi /2)$, not discussed in \cite{KW}, show that (\ref{littleh}%
) indeed contains nonlocal bulk interactions. Namely consider the
non-Hermitian Hamiltonian%
\begin{equation}
H_{g}=-\frac{1}{2}\sum_{m=1}^{M-1}\left[ \sigma _{m}^{x}\sigma
_{m+1}^{x}+\sigma _{m}^{y}\sigma _{m+1}^{y}+ig(\sigma _{m}^{z}-\sigma
_{m+1}^{z})\right] ,\quad 0<g<1,  \label{Hg}
\end{equation}%
then one can show for small $g$ its spectral equivalence with the Hermitian
Hamiltonian%
\begin{equation}
h_{g^{2}}=-\sum_{n>0}\sum_{x=1}^{M-n}p_{x}^{(n)}(g^{2})\left[ c_{x}^{\ast
}c_{x+n}-c_{x}c_{x+n}^{\ast }\right] ,  \label{hg}
\end{equation}%
where $c_{x}^{\ast },c_{x}$ are fermionic creation and annihilation
operators at lattice site $x$. The hopping probability between a site $x$
and its $n^{\text{th}}$ neighbour is encoded in the real coefficients $%
p_{x}^{(n)}$. Up to order $g^{8}$ one finds that the nonvanishing
contributions are%
\begin{eqnarray}
p_{x}^{(1)} &=&1-\frac{128g^{2}+8g^{4}+g^{6}}{512}(\delta _{x,1}+\delta
_{x,M-1}),  \notag \\
&&-\frac{8g^{4}+3g^{6}}{512}(\delta _{x,2}+\delta _{x,M-2})+\frac{g^{6}}{256}%
(\delta _{x,3}+\delta _{x,M-3})+O(g^{8}),  \notag \\
p_{x}^{(3)} &=&\frac{20g^{4}+3g^{6}}{256}~(\delta _{x,1}+\delta _{x,M-3})+%
\frac{5g^{6}}{512}(\delta _{x,2}+\delta _{x,M-4})+O(g^{8})  \notag \\
p_{x}^{(5)} &=&-\frac{23g^{6}}{512}(\delta _{x,1}+\delta _{x,M-5})+O(g^{8}),
\end{eqnarray}%
A more detailed discussion of this model, which warrants an investigation in
its own right, will be presented elsewhere \cite{CK}.

The close investigation of the regime (ii), $N<r<\infty ,$ might shed light
on how to perform similar GNS constructions for other sections of the unit
circle. Such a purely algebraic formulation is desirable in order to apply
the results to a wider range of Temperley-Lieb type models, see e.g.\cite%
{Kulish} and references therein. For instance, by showing the more
restrictive identity $\eta E_{i}=E_{i}^{\ast }\eta $ for the Temperley-Lieb
algebra generators the results immediately generalize also to transfer
matrices. Namely, it easily follows from the results presented in \cite{KW}
and this article that we have constructed a unitary representation of the
Hecke algebra%
\begin{equation}
\eta B_{i}=(B_{i}^{-1})^{\ast }\eta ,\qquad B_{i}=q^{-1}+E_{i}\ .
\end{equation}%
The latter form the basic building blocks for transfer matrices, as for
instance the (non-symmetric) double row transfer matrices considered in
connection with lattice systems associated to logarithmic minimal models in 
\cite{PRZ06}. With regard to these applications it is natural to ask if the
change of the inner product will effect the description in terms of
logarithmic conformal field theory when $N\rightarrow \infty $ and how this
connects to possibly non-local interactions in the bulk. It is planned to
address these questions in future work.

Finally, it is worth noting that we discussed in this article the problem of
a non-Hermitian quantum Hamiltonian in the language of $C^{\ast }\,$%
-algebras. Instead of using the concepts of quasi-Hermiticity or $PT$%
-symmetry we have considered an associated $C^{\ast }$-algebra of the given
quantum mechanical system and investigated the existence of positive linear
functionals. The latter have then be used to construct via the GNS approach
an inner product with respect to which the Hamiltonian in question is
Hermitian. The formulation of quantum mechanics in the language of $C^{\ast }
$-algebras is not new, but the novel aspect in this article is the
presentation of an explicit example where this formulation can be connected
with quasi-Hermiticity and $PT$-symmetry.{\small \medskip }

\noindent \textbf{Acknowledgments}. The author would like to thank Catharina
Stroppel and Robert Weston for many helpful discussions. C.K. is financially
supported by a University Research Fellowship of the Royal Society.


\begin{thebibliography}{99}
\bibitem{KW} C. Korff and R.~Weston. 
\newblock {PT symmetry on the lattice:
the quantum group invariant XXZ spin-chain}. \newblock {\em J. Phys.},
A40:8845--8872, 2007.

\bibitem{Alc87} F.~C. Alcaraz, M.~N. Barber, M.~T. Batchelor, R.~J. Baxter,
and G.~R.~W. Quispel. 
\newblock {Surface exponents of the quantum XXZ, Ashkin-Teller and Potts
  models}. \newblock {\em J. Phys.}, A20:6397--6409, 1987.

\bibitem{PS90} V.~Pasquier and H.~Saleur. 
\newblock {Common Structures Between finite Systems and Conformal field
  Theories Through Quantum Groups}. \newblock {\em Nucl. Phys.},
B330:523--556, 1990.

\bibitem{SGH92} F.~G.~Scholtz, H.~B.~Geyer, and F.~J.~W.~Hahne. \newblock
{Quasi-Hermitian operators in quantum mechanics and the variational principle%
}. \newblock {\em Ann. of Phys.}, 213:74--101, 1992.

\bibitem{Most04} A.~Mostafazadeh. 
\newblock {Physical Aspects of Pseudo-Hermitian and $PT$-Symmetric Quantum
  Mechanics}. \newblock {\em J. Phys. A} 37:11645--11680, 2004.

\bibitem{RS90} N.~Reshetikhin and F.~Smirnov. 
\newblock {Hidden quantum group symmetry and integrable perturbations of
  conformal field theories}. \newblock {\em Comm. Math. Phys.}, 131:157,
1990.

\bibitem{Bender07} C.~M. Bender. 
\newblock {Making sense of non-Hermitian
Hamiltonians}, 2007. \newblock hep-th/0703096.

\bibitem{CFMFAF} C. Figueira de Morisson Faria and A. Fring, 
\newblock{Non-Hermitian Hamiltonians with real eigenvalues coupled to electric fields:
from the time-independent to the time dependent quantum mechanical
formulation}. \newblock {\em Laser Physics}, 17:424-437, 2007.

\bibitem{Jim86b} M.~Jimbo. 
\newblock {A q-analogue of $U(gl(N + 1))$, Hecke algebra, and the Yang-Baxter
  equation}. \newblock {\em Lett. Math. Phys.}, 11:247--252, 1986.

\bibitem{Jo83} V.~Jones. \newblock {Index for Subfactors}. 
\newblock {\em
{Invent. math.}}, 72:1--25, 1983.

\bibitem{We88} H.~Wenzl. 
\newblock {Hecke Algebras of type $A_n$ and
subfactors}. \newblock {\em Invent. math.}, 92:349--383, 1988; 
\newblock {Quantum Groups and Subfactors of Type B,
C and D}. \newblock {\em Commun. Math. Phys.}, 133:383--432, 1990.

\bibitem{PPM91} Paul Martin. 
\newblock {\em {Potts Models And Related
Problems In Statistical Mechanics}}. \newblock World Scientific, 1991.

\bibitem{Kauf87} L.~H. Kauffman. \newblock State models and the {J}ones
polynomial. \newblock {\em Topology}, 26(3):395--407, 1987.

\bibitem{Jo99} V.~Jones. \newblock {Planar Algebras, I}.
arXiv:math/9909027v1; \newblock {The Jones Polynomial}.
http://math.berkeley.edu/\symbol{126}vfr/

\bibitem{CK} C.~Korff, in preparation

\bibitem{Kulish} P.~P.~Kulish. 
\newblock{On spin systems related to the
Temperley--Lieb algebra}. \newblock {\em J. Phys. A: Math. Gen.},
36:L489-L493, 2003.

\bibitem{PRZ06} P.~A.~Pearce, J.~Rasmussen, and J.-B.~Zuber. %
\newblock{Logarithmic minimal models}. 
\newblock {\em J. Stat. Mech.: Theory
and Experiment}, P11017, 2006.
\end{thebibliography}
\end{document}